\documentclass[preprintnumbers,amsmath,amssymb]{revtex4}
\usepackage{epsfig}
\usepackage{color}
\usepackage{dsfont}
\usepackage[colorlinks,urlcolor=blue,linkcolor=blue,citecolor=red]{hyperref}
\begin{document}
\setlength{\voffset}{1.0cm}
\title{Phase structure of the 1+1 dimensional Nambu--Jona-Lasinio model with isospin}
\author{Michael Thies\footnote{michael.thies@gravity.fau.de}}
\affiliation{Institut f\"ur  Theoretische Physik, Universit\"at Erlangen-N\"urnberg, D-91058, Erlangen, Germany}
\date{\today}

\begin{abstract}
The phase diagram of the two-dimensional Nambu--Jona-Lasinio model with isospin is explored in the large $N_c$ limit with semiclassical methods. We consider finite temperature and 
include chemical potentials for all conserved charges. In the chiral limit, a full analytical solution is presented, expressed in terms of known results for the single-flavor Gross-Neveu and
Nambu--Jona-Lasinio models. A novel crystalline structure appears and is shown explicitly to be thermodynamically more stable than the homogeneous phase at zero temperature.
If we include a bare fermion mass, the problem reduces again to solved problems in one-flavor models provided that either the fermionic or the isospin chemical potentials vanish.
In the general case, a stability analysis is used to construct the perturbative phase boundary between homogeneous and inhomogeneous phases. This is sufficient to get a good
overview of the complete phase diagram. Missing nonperturbative phase boundaries requiring a full numerical Hartree-Fock calculation will be presented in future work.
\end{abstract}

\maketitle

\section{Introduction}
\label{sect1}

The two most widely studied versions of the Gross-Neveu (GN) model in 1+1 dimensions are the original one \cite{L1} with Lagrangian
\begin{equation}
{\cal L}_{\rm GN} = \bar{\psi} ( i \partial \!\!\!/ - m_0)  \psi + \frac{g^2}{2}  (\bar{\psi} \psi)^2 
\label{1.1}
\end{equation}
and the chiral GN or 1+1 dimensional Nambu--Jona-Lasinio (NJL) model \cite{L2},
\begin{equation}
{\cal L}_{\rm NJL} = \bar{\psi} ( i \partial \!\!\!/ - m_0)  \psi + \frac{g^2}{2} \left[ (\bar{\psi} \psi)^2+ (\bar{\psi} i \gamma_5 \psi)^2 \right] .
\label{1.2}
\end{equation}
In the massless limit ($m_0=0$), ${\cal L}_{\rm GN}$ has a discrete  Z(2)$_L \times$Z(2)$_R$ chiral symmetry, promoted to a continuous U(1)$_L \times$U(1)$_R$
chiral symmetry in ${\cal L}_{\rm NJL}$.
The Dirac fermions come in $N_c$ ``colors" (a name for flavor in this context), and such models are  typically solved in the large $N_c$ limit \cite{L3} with semiclassical methods. 
Our focus here will be on equilibrium thermodynamics and the phase diagrams. The latter have been established some time ago for both the massless \cite{L4,L5} and the massive \cite{L6,L7}
variants of models (\ref{1.1}) and (\ref{1.2}). The biggest surprise was probably the emergence of inhomogeneous phases. While this had been overlooked at first \cite{L8,L9}, it could
have been anticipated on the basis of the Peierls instability \cite{L10}, ubiquitous in one-dimensional condensed matter systems.

The most popular NJL model in 3+1 dimensions, an effective field theory for strong interaction physics, has isospin in addition to color \cite{L11}. 
The corresponding Lagrangian reads
\begin{equation}
{\cal L}_{\rm isoNJL} = \bar{\psi} ( i \partial \!\!\!/ - m_0)  \psi + \frac{G^2}{2} \left[ (\bar{\psi} \psi)^2+ (\bar{\psi} i \gamma_5 \vec{\tau}\psi)^2 \right]  
\label{1.3}
\end{equation}
and features SU(2)$_L \times$SU(2)$_R$ chiral symmetry for $m_0=0$. 
Isospin is obviously of great importance for phenomenological applications, but this variant has received less attention in 1+1 dimensions.
Interest has grown only recently. In spite of several investigations addressing the thermodynamics of the model, the understanding of the phase diagram is 
still incomplete. To date we have mostly information about the chiral limit.  
Using an (unbiased) finite mode approach, Heinz {\em et al.} \cite{L12} find numerically that the phase diagram of the massless isoNJL model in the ($\mu,T$)-plane 
is identical to that of the GN model. An explanation of this remarkable coincidence and a sketch of a possible phase diagram including an isospin chemical potential was outlined in Ref.~\cite{L13},
but without detailed calculations. The Moscow group \cite{L14,L15} has presented several variational calculations of the phase diagram including isospin and
chiral imbalance, both with homogeneous and inhomogeneous mean fields. They emphasize the phenomenon of charged pion condensation and a certain duality. 
These results are complementary to those of Ref.~\cite{L12}, but have not yet led to a definite picture of the phase diagram.
The phase diagram of the massive isoNJL model has also been addressed within several variational calculations in recent years \cite{L16,L17,L18}.

Evidently, the isoNJL model is the most complicated one of the GN family (\ref{1.1})--(\ref{1.3}). Due to the possible presence of up to three distinct chemical potentials
and a bare mass, mapping out the full phase diagram is quite a challenge. Nevertheless we believe that the toolbox developed in the past for solving models (\ref{1.1}) and (\ref{1.2})
should contain everything necessary for determining the complete phase diagram of the isoNJL model as well, using a combination of numerical and analytical methods. 
It is the purpose of this paper to fill the gap in our understanding of the phase structure of GN type models by constructing this missing phase diagram. 

Throughout this paper, we shall frequently have to refer to one of the three GN type models. For the sake of simplicity, we call model (\ref{1.1}) GN model,
model (\ref{1.2}) NJL model and model (\ref{1.3}) isoNJL model (NJL model with isospin). Since we always work in 1+1 dimensions, we refrain
from using subscripts referring to the number of dimensions as, e.g., in NJL$_2$ model.

This paper is organized as follows. Sections.~\ref{sect2} and \ref{sect3} deal with the chiral limit of the isoNJL model, whereas Sec.~\ref{sect4} addresses the massive
model. More specifically, in Sec.~\ref{sect2} we identify special cases where the phase diagram of the isoNJL model can be rigorously reduced to that 
of the GN or the NJL model. Section~\ref{sect3} presents the full phase diagram in the chiral limit, derived analytically. A novel kind of crystal phase is 
identified and illustrated. Section~\ref{sect4} is dedicated to the massive isoNJL model. As in Sec.~\ref{sect2}, the phase boundaries can be inferred from those 
of the massive GN and NJL models in certain cases. We then construct the perturbative phase boundary separating the crystal from the homogeneous phase 
for a wide range of chemical potentials and bare masses. In the concluding section, Sec.~\ref{sect5}, we summarize our findings and point out areas where further
numerical work is needed.

\section{Chiral limit -- Reduction of the isoNJL model to GN and NJL models}
\label{sect2}

The massless isoNJL model has the standard U(1) vector symmetry (conservation of fermion number) as well as SU(2)$_L \times$SU(2)$_R$ chiral symmetry (conservation of 
isospin and axial isospin charges). Consequently we shall be interested in the phase diagram of the isoNJL model with three different chemical potentials.
Its Lagrangian reads
\begin{equation}
{\cal L} = \bar{\psi} i \partial \!\!\!/ \psi + \frac{G^2}{2} \left[ (\bar{\psi} \psi)^2+ (\bar{\psi} i \gamma_5 \vec{\tau}\psi)^2 \right] + \mu \psi^{\dagger} \psi
+ \nu \psi^{\dagger} \tau_3 \psi + \nu_5 \psi^{\dagger} \tau_3 \gamma_5 \psi .
\label{2.1}
\end{equation}
In the large $N_c$ limit, the thermal Hartree-Fock (HF) approximation is adequate. It can be formulated concisely as single particle Dirac-HF equation
\begin{equation}
\left(-i \gamma_5 \partial_x + \gamma^0 S + i \gamma^1 \tau_a P_a - \mu - \nu \tau_3 - \nu_5 \gamma_5  \tau_3  \right)\psi = \omega \psi
\label{2.2}
\end{equation}
supplemented by the self-consistency conditions for scalar and pseudoscalar mean fields,
\begin{eqnarray}
S & = & -  G^2 \langle \bar{\psi} \psi \rangle ,
\nonumber  \\
P_a & = & - G^2 \langle \bar{\psi} i \gamma_5 \tau_a \psi \rangle .
\label{2.3}
\end{eqnarray}
The brackets denote either thermal averages ($T\neq 0$) or ground state averages ($T=0$). Once the HF problem is solved, one
can compute the grand canonical potential and all thermodynamic observables by standard methods.

We choose the following representation of the Dirac matrices,
\begin{equation}
\gamma^0 = \sigma_1, \quad \gamma^1 = i \sigma_2, \quad \gamma_5 = \gamma^0 \gamma^1 = - \sigma_3 ,
\label{2.4}
\end{equation} 
and denote 4-component spinors/isospinors as
\begin{equation}
\left( \begin{array}{c} \Psi_{1,1} \\ \Psi_{1,2} \\ \Psi_{2,1} \\ \Psi_{2,2} \end{array} \right) = 
\left( \begin{array}{c} \Psi_{L,\uparrow} \\ \Psi_{L,\downarrow} \\ \Psi_{R,\uparrow} \\ \Psi_{R,\downarrow} \end{array} \right) .
\label{2.5}
\end{equation}
Thus the two indices on the components $\Psi_{i,j}$ label chirality ($L,R$) and the 3rd component of isospin (up,down), respectively. The (grand canonical) HF Hamiltonian
in this representation assumes the form
\begin{equation}
H=
\left( \begin{array}{cccc} i \partial_x - \mu - \nu + \nu_5 & 0 & {\cal D}^* & {\cal C}^* \\ 0 & i \partial_x - \mu + \nu -\nu_5 & - {\cal C} & {\cal D} \\
{\cal D} & - {\cal C}^* & - i \partial_x - \mu - \nu - \nu_5  & 0 \\ {\cal C} & {\cal D}^* & 0 & - i \partial_x - \mu + \nu + \nu_5  \end{array} \right) 
\label{2.6}
\end{equation}
with two complex potentials defined as 
\begin{equation}
{\cal D} = S-i P_3, \quad {\cal C} = P_2-iP_1 .
\label{2.7}
\end{equation}
Using Eqs.~(\ref{2.3}) and (\ref{2.7}), the self-consistency conditions become
\begin{eqnarray}
{\cal D} & = & - 2 N_c G^2 \sum \left(\Psi_{1,1}^* \Psi_{2,1} + \Psi_{2,2}^* \Psi_{1,2} \right) n_{\rm occ} ,
\nonumber \\
{\cal C} & = &  - 2 N_c G^2  \sum \left(\Psi_{1,1}^* \Psi_{2,2} - \Psi_{2,1}^* \Psi_{1,2} \right) n_{\rm occ} ,
\label{2.8}
\end{eqnarray}
with the occupation numbers
\begin{equation}
n_{\rm occ} = \frac{1}{e^{\beta \omega} + 1} , \quad \beta=\frac{1}{T} \ {\rm \ for \ } T \neq 0,  \quad \quad n_{\rm occ} = \theta(- \omega) \ {\rm \ for \ } T=0 .
\label{2.9}
\end{equation}
As pointed out and discussed extensively by the Moscow group \cite{L14,L15}, the Hamiltonian (\ref{2.6}) exhibits an interesting duality. We may phrase it in the following way: Consider  
a particular element of the original, global SU(2)$\times$SU(2) chiral symmetry group of the isoNJL model,
\begin{equation}
U_{\rm dual} = i \tau_3 P_L + i \tau_1 P_R, \quad P_{R,L} = \frac{1\pm \gamma_5}{2}.
\label{2.10}
\end{equation}  
It interchanges ${\cal D}$ and ${\cal C}$, as well as $\nu$ and $-\nu_5$,
\begin{equation}
U_{\rm dual} H({\cal D},{\cal C}, \mu, \nu, \nu_5) U_{\rm dual}^{\dagger} = H({\cal C}, {\cal D}, \mu, -\nu_5, -\nu).
\label{2.11}
\end{equation}
In the absence of isospin chemical potentials, this would be physically irrelevant, simply amounting to a
different choice of the vacuum on the SU(2) manifold. In the presence of isospin chemical potentials,
it may indeed be relevant since the global SU(2)$\times$SU(2) symmetry is broken down explicitly to U(1)$\times$U(1).   

There are two obvious cases where $H$ assumes a block diagonal form: ${\cal C}=0$ (only ``scalar" and ``neutral pion" condensates ${\cal D}$), and ${\cal D}=0$
(only ``charged pion" condensates ${\cal C}$). We shall stick to these special cases throughout this paper and see how far we can get. 
For homogeneous condensates, Khunjua {\em et al.} \cite{L15} have verified numerically that all extrema of the effective potential belong to this class of solutions.
For inhomogeneous condensates, this still remains to be checked. Moreover, we only consider the neutral condensate case (${\cal C}=0$). Below we argue that 
this is general enough for our purpose. Upon setting ${\cal C}=0$, the
HF equation decouples into two HF equations of the one-flavor NJL model, one for each isospin component,
\begin{eqnarray}
\left( \begin{array}{cc} i \partial_x-\mu-\nu+\nu_5 & {\cal D}^* \\ {\cal D}  & - i \partial_x - \mu - \nu - \nu_5 \end{array} \right) 
\left( \begin{array}{c} \Psi_{1,1} \\ \Psi_{2,1} \end{array} \right) & = &  \omega \left( \begin{array}{c} \Psi_{1,1} \\ \Psi_{2,1} \end{array} \right) ,
\nonumber \\
\left( \begin{array}{cc} i \partial_x-\mu+\nu-\nu_5 & {\cal D} \\ {\cal D}^*  & - i \partial_x - \mu + \nu + \nu_5 \end{array} \right) 
\left( \begin{array}{c} \Psi_{1,2} \\ \Psi_{2,2} \end{array} \right) & = &  \omega \left( \begin{array}{c} \Psi_{1,2} \\ \Psi_{2,2} \end{array} \right) .
\label{2.12}
\end{eqnarray}
Eqs.~(\ref{2.12}) belong to  NJL models with different chemical and axial chemical potentials, but complex conjugate mean fields.
This prevents us in general from using the known solution of the NJL model to solve the isoNJL model. The formal reason why the problem does not 
really separate are the self-consistency conditions, different in the one- and two-flavor cases.
If the two NJL models of (\ref{2.12}) would be really independent, the first line
would correspond to the one-flavor theory with chemical potential $\mu+\nu$, axial chemical potential $ \nu_5$, mean field ${\cal D}$
and be supplemented by the self-consistency condition
\begin{equation}
{\cal D} = - 2 N_c g^2 \sum \left( \Psi_{1,1}^* \Psi_{2,1} \right) n_{\rm occ} .
\label{2.13}
\end{equation}
The 2nd line of (\ref{2.12}) would correspond to the same model but with chemical potential $\mu-\nu$, axial chemical potential $-\nu_5$, mean field ${\cal D}^*$ and
self-consistency condition
\begin{equation}
{\cal D}^* = - 2 N_c g^2 \sum \left( \Psi_{1,2}^* \Psi_{2,2}  \right) n_{\rm occ} .
\label{2.14}
\end{equation}
Conditions (\ref{2.13}) and (\ref{2.14}) are in general incompatible, since different chemical potentials would require mean fields not related by complex conjugation.
Anyway, neither condition (\ref{2.13}) nor condition (\ref{2.14}) holds here. We rather have to  make sure that the isoNJL self-consistency conditions (\ref{2.8}) for the
special case ${\cal C}=0$ are satisfied, i.e., 
\begin{eqnarray}
{\cal D} & = & - 2 N_c G^2 \sum \left(\Psi_{1,1}^* \Psi_{2,1} + \Psi_{2,2}^* \Psi_{1,2} \right) n_{\rm occ} ,
\nonumber \\
0 & = &  - 2 N_c G^2  \sum \left(\Psi_{1,1}^* \Psi_{2,2} - \Psi_{2,1}^* \Psi_{1,2} \right) n_{\rm occ} .
\label{2.15}
\end{eqnarray}
Thus one cannot solve the isoNJL model simply with the solution of the NJL model in general.
There are two important special cases though where one gets away with the knowledge of the one-flavor model:
keeping only the fermion chemical potential $\mu$, or keeping only isospin and chiral isospin chemical potentials $\nu, \nu_5$. 
We shall return to the general case in the next section.
\vskip 0.2cm
{\bf Case I: \ \ \ $\nu=\nu_5=0, \ \mu \neq 0$}
\vskip 0.2cm
Here we are dealing with hot and dense isospin-symmetric matter. Equation~(\ref{2.12}) reduces to 
\begin{eqnarray}
\left( \begin{array}{cc} i \partial_x-\mu & {\cal D}^* \\ {\cal D}  & - i \partial_x - \mu \end{array} \right) 
\left( \begin{array}{c} \Psi_{1,1} \\ \Psi_{2,1} \end{array} \right) & = &  \omega \left( \begin{array}{c} \Psi_{1,1} \\ \Psi_{2,1} \end{array} \right) ,
\nonumber \\
\left( \begin{array}{cc} i \partial_x-\mu  & {\cal D} \\ {\cal D}^*  & - i \partial_x - \mu  \end{array} \right) 
\left( \begin{array}{c} \Psi_{1,2} \\ \Psi_{2,2} \end{array} \right) & = &  \omega \left( \begin{array}{c} \Psi_{1,2} \\ \Psi_{2,2} \end{array} \right) .
\label{2.16}
\end{eqnarray}
Since now the chemical potentials are the same, the mean fields must also agree: ${\cal D}={\cal D}^* =S$.
Hence the two equations reduce to identical equations for the GN model with discrete chiral symmetry at chemical potential $\mu$. 
There we know the answer. Depending on the location in the ($\mu,T$)-plane, $S_{\rm GN}$ is 0 (symmetric phase), constant 
(homogeneous phase) or has the form of a kink crystal (see Fig.~\ref{fig1} below).
Denoting the GN spinor components by $\psi_{1,2}$, the one-flavor self-consistency condition reads
\begin{equation}
S_{\rm GN} = - N_c g^2 \sum \left( \psi_1^* \psi_2 + \psi_2^* \psi_1 \right) n_{\rm occ}.
\label{2.17}
\end{equation}
Note that this is only the real part of the NJL self-consistency condition. 
From this we can construct an exact solution of the isoNJL model. Let us choose the following spinors in the two decoupled isospin channels,
\begin{equation}
 \left( \begin{array}{c} \Psi_{1,1} \\ \Psi_{1,2} \\ \Psi_{2,1} \\ \Psi_{2,2} \end{array} \right)_{\uparrow} = \left( \begin{array}{c} \psi_1 \\ 0 \\ \psi_2 \\ 0 \end{array} \right), 
\quad \left( \begin{array}{c} \Psi_{1,1} \\ \Psi_{1,2} \\ \Psi_{2,1} \\ \Psi_{2,2} \end{array} \right)_{\downarrow} =
 \left( \begin{array}{c} 0 \\ \psi_1 \\ 0 \\ \psi_2  \end{array} \right) .
\label{2.18}
\end{equation}
Equations~(\ref{2.16}) are satisfied by construction. The first isoNJL self-consistency condition (\ref{2.15}) becomes
\begin{equation}
{\cal D}  =  - 2 N_c G^2 \sum \left(\psi_{1}^* \psi_{2} + \psi_{2}^* \psi_{1} \right) n_{\rm occ}
\label{2.19}
\end{equation}
where the two terms on the right-hand side arise from isospin up and isospin down contributions, respectively. This matches the single flavor GN model (\ref{2.17}),
provided we identify $2 N_c G^2$ with $N_c g^2$. But this is precisely what the gap equation tells us if we use the same ultraviolet (UV) cutoff $\Lambda/2$ in the isoNJL and GN models.
We remind the reader of the following vacuum gap equations (using units where $m=1$)
\begin{eqnarray}
\frac{\pi}{N_c g^2} &=& \ln \Lambda \quad {\rm (GN\ or\ NJL)} ,
\nonumber \\
\frac{\pi}{2 N_c G^2} & = & \ln \Lambda \quad {\rm (isoNJL)} .
\label{2.20}
\end{eqnarray}
The difference between the cases with and without isospin reflects the fact that the total number of flavors has increased by a factor of 2.
The 2nd line of the isoNJL self-consistency condition (\ref{2.15}) is trivially satisfied for the spinors (\ref{2.18}). Hence the isoNJL model at finite $\mu,T$ 
can be solved using the known solution of the GN model,
\begin{equation}
{\cal D}_{\rm isoNJL}(\mu,\nu=0,\nu_5=0,T) = S_{\rm GN}(\mu,T) .
\label{2.21}
\end{equation} 
This fully confirms and explains the numerical findings of Heinz {\em et al.} \cite{L12}.
The thermodynamic potentials are then related as follows,
\begin{equation}
\left. \frac{{\cal V}_{\rm eff}(\mu,\nu=\nu_5=0,T)}{2 N_c} \right|_{\rm isoNJL} = \left. \frac{{\cal V}_{\rm eff}(\mu,T)}{N_c} \right|_{\rm GN} .
\label{2.22}
\end{equation}
\vskip 0.2cm
{\bf Case II: \ \ \ $\mu=0, \ \nu \neq 0 ,\ \nu_5 \neq 0$}
\vskip 0.2cm
This is the case of finite isospin density with vanishing fermion density. In quantum chromodynamics with $N_c=3$, one would associate a pion condensate with such a phase.
Here, due to the large $N_c$ limit, we prefer to think of a system with the same density of up quarks and down antiquarks (or vice versa). 
Equation~(\ref{2.12}) reduces to 
\begin{eqnarray}
\left( \begin{array}{cc} i \partial_x- \nu+\nu_5 & {\cal D}^* \\ {\cal D}  & - i \partial_x  - \nu - \nu_5 \end{array} \right) 
\left( \begin{array}{c} \Psi_{1,1} \\ \Psi_{2,1} \end{array} \right) & = &  \omega \left( \begin{array}{c} \Psi_{1,1} \\ \Psi_{2,1} \end{array} \right) ,
\nonumber \\
\left( \begin{array}{cc} i \partial_x +\nu-\nu_5 & {\cal D} \\ {\cal D}^*  & - i \partial_x  + \nu + \nu_5 \end{array} \right) 
\left( \begin{array}{c} \Psi_{1,2} \\ \Psi_{2,2} \end{array} \right) & = &  \omega \left( \begin{array}{c} \Psi_{1,2} \\ \Psi_{2,2} \end{array} \right) .
\label{2.23}
\end{eqnarray}
The first line is the HF equation of the single flavor NJL model with chemical potential $\nu$ and axial chemical potential $\nu_5$. Its solution is known
as chiral spiral (spinor components $\psi_{1,2}$, $\Delta_{\rm NJL}=S-iP$),
\begin{equation}
{\Delta}_{\rm NJL}= m(T) e^{2i\nu x} = - 2 N_c g^2 \sum \psi_1^*\psi_2 n_{\rm occ} .
\label{2.24}
\end{equation}
Here $m(T)$ is the thermal mass of the fermions at $\mu=0$, vanishing for  $T \ge T_{\rm crit}$. The chiral isospin chemical potential $\nu_5$ does not show up in the mean field but in some observables,
as it induces an asymmetric UV cutoff \cite{L19}.  
The second line of (\ref{2.23}) is the corresponding solution with complex conjugate mean field, opposite chemical potentials ($-\nu,-\nu_5$) and spinor components $\phi_{1,2}$,
\begin{equation}
{\Delta}_{\rm NJL}^*= m(T) e^{-2i\nu x} = - 2 N_c g^2 \sum \phi_1^*\phi_2 n_{\rm occ} 
\label{2.25}
\end{equation}
or, equivalently,
\begin{equation}
{\Delta}_{\rm NJL}= m(T) e^{2i\nu x} = - 2 N_c g^2 \sum \phi_2^*\phi_1 n_{\rm occ} .
\label{2.26}
\end{equation}
Choosing the isospin up and down spinors
\begin{equation}
 \left( \begin{array}{c} \Psi_{1,1} \\ \Psi_{1,2} \\ \Psi_{2,1} \\ \Psi_{2,2} \end{array} \right)_{\uparrow} = \left( \begin{array}{c} \psi_1 \\ 0 \\ \psi_2 \\ 0 \end{array} \right), 
\quad \left( \begin{array}{c} \Psi_{1,1} \\ \Psi_{1,2} \\ \Psi_{2,1} \\ \Psi_{2,2} \end{array} \right)_{\downarrow} =
 \left( \begin{array}{c} 0 \\ \phi_1 \\ 0 \\ \phi_2  \end{array} \right) 
\label{2.27}
\end{equation}
as solutions of the isoNJL HF equation (\ref{2.23}), the self-consistency condition for the isoNJL model reads [see first line of Eq.~(\ref{2.15})]
\begin{equation}
{\cal D} = - 2 N_c G^2 \sum \left( \psi_1^* \psi_2 + \phi_2^* \phi_1\right) n_{\rm occ} .
\label{2.28}
\end{equation}
This matches again the single-flavor self-consistency relations (\ref{2.24}--\ref{2.26}) for $g^2=2G^2$.  
The 2nd line in (\ref{2.15}) is again trivially fulfilled. Hence we have found a HF solution of the isoNJL model in the case $\mu=0$ in terms of  
solutions of the NJL model. The phase diagram in the ($\nu,T$) plane is the same as the NJL phase diagram in the ($\mu,T$) plane and also 
shown in Fig.~\ref{fig1}. 
For the thermodynamic potential, we obtain
\begin{eqnarray}
\left. \frac{{\cal V}_{\rm eff}(\mu=0,\nu,\nu_5,T)}{2 N_c} \right|_{\rm isoNJL} & = &  \left. \frac{{\cal V}_{\rm eff}(\mu=\nu,\mu_5=\nu_5,T)}{N_c} \right|_{\rm NJL}
\nonumber \\
& = &  \left. \frac{{\cal V}_{\rm eff}(\mu=0,T)}{N_c} \right|_{\rm GN} - \frac{\nu^2+\nu_5^2}{2 \pi}.
\label{2.29}
\end{eqnarray}
The last line is taken over from Ref.~\cite{L19} where the NJL model with chiral imbalance has been discussed.

Summarizing cases I and II, we learn that depending on the choice of chemical potentials, the isoNJL model can mimick two well-known phase diagrams:
those of the GN and the NJL model, prime examples of phase diagrams with inhomogeneous phases. The coincidence with the GN model in the ($\mu,T$)-plane
was already discovered numerically in Ref.~\cite{L12}. In view of our goal to construct the full phase diagrams of the (massless and massive) isoNJL models
in the whole parameter space (temperature, chemical potentials), these analytical findings on the boundary give very useful constraints.
Figure~\ref{fig1} summarizes what we have learned so far about the phase diagram of the massless isoNJL model.
\begin{figure}
\begin{center}
\epsfig{file=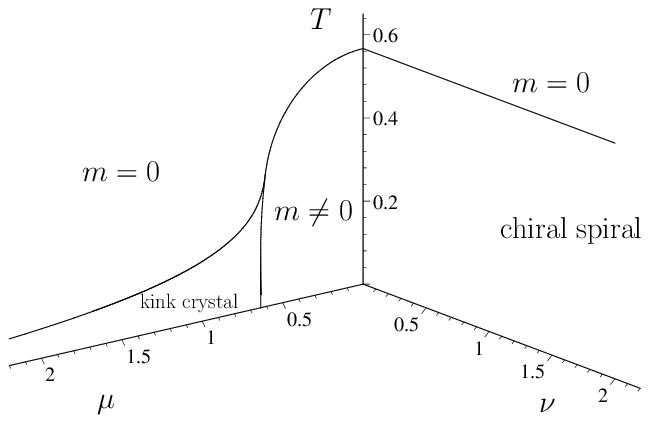,width=8cm,angle=0}
\caption{Phase diagram of the massless isoNJL model in the ($\mu,T$) and ($\nu,T$) planes, coinciding with phase diagrams of the one-flavor GN and NJL models,
respectively. In the latter case, $\mu$ has to be replaced by $\nu$.}
\label{fig1}
\end{center}
\end{figure}

Up to this point, we have tacitly assumed that ${\cal C}=0$. By applying the duality transformation (\ref{2.10}), we can swap ${\cal D}$ with ${\cal C}$ and $\nu$ with $-\nu_5$.
In the first special case ($\nu=0,\nu_5=0$), this gives nothing new, although the mean field looks quite different. This is due to the fact that, in the case of spontaneous symmetry breaking, 
picking a particular point on the vacuum manifold cannot have any observable consequences.
The choice ${\cal C}=0$ is singled out if one approaches the chiral limit from the massive theory side ($m_0 \to 0$), so that
we would rather stick to it. In the second special case ($\mu=0$), the value of the thermodynamic potential also does not change
[it is symmetric under $\nu \leftrightarrow -\nu_5$, see (\ref{2.29})]. However the $\nu$-axis in Fig.~\ref{fig1} would become the $\nu_5$ axis, and the 
breakdown of translational invariance would have to be attributed to $\nu_5$ rather than $\nu$. We believe that this reflects the freedom of  
choosing different isospin frames for left- and right-handed fermions in the isoNJL model. No such ambiguity exists in the one-flavor NJL model 
where only $\mu$ can induce crystallization, not $\mu_5$. 
If we invoke again the limit $m_0 \to 0$ to define the chiral limit, the choice ${\cal C}=0$ and Fig.~\ref{fig1} with the $\nu$ axis as shown are singled out. 

\section{Chiral limit -- Full phase diagram of the isoNJL model}
\label{sect3}

In the preceding section we have considered special cases where one or two of the chemical potentials ($\mu, \nu, \nu_5$) vanish. Then the isoNJL model problem could be reduced exactly to the simpler one-flavor
NJL and GN models. This is no longer the case once we switch on all three chemical potentials. Nevertheless, the experience from the simpler models can guide us to an exact HF solution.
At $\mu=0$, we can generate the isospin chemical potentials $\nu$ and $\nu_5$ by a (local) chiral transformation by means of the axial anomaly, see \cite{L13,L19}. Since this is an UV effect,
the same trick can be used at  $\mu \neq 0$ and finite temperature. We therefore start from the case $\nu=\nu_5=0$ where the HF equation reads (see special case I above) 
\begin{equation}
\left(-i \gamma_5 \partial_x + \gamma^0 S_{\rm GN}  - \mu \right)\Psi = \omega \Psi.
\label{3.1}
\end{equation}
At this level, there is no isospin dependence. The self-consistent HF potential $S_{\rm GN}$ is real and may be 0, a dynamical mass or a periodic function of $x$,  
depending on where one is in the ($\mu,T$)-plane (see Fig.~\ref{fig1}).
Let us apply the unitary transformation
\begin{equation}
\Psi = U\Phi, \quad U = e^{-i x (\nu_5+ \nu  \gamma_5) \tau_3}.
\label{3.2}
\end{equation}
The HF equation goes over into
\begin{equation}
\left(-i \gamma_5 \partial_x + U^{\dagger}\gamma^0 U  S_{\rm GN}  - \mu  - i \left[ U^{\dagger} \gamma_5 \partial_x U \right]  \right)\Phi = \omega \Phi.
\label{3.3}
\end{equation}
Upon plugging in $U$ from (\ref{3.2}), we find
\begin{equation}
\left(-i \gamma_5 \partial_x + \gamma^0 S + i \gamma^1 \tau_3 P_3 - \mu - \nu \tau_3 - \nu_5 \tau_3 \gamma_5 \right)\Phi = \omega \Phi
\label{3.4}
\end{equation}
with
\begin{equation}
S= S_{\rm GN}\cos 2\nu x, \quad P_3 = - S_{\rm GN} \sin 2\nu x, \quad {\cal D} = S-iP_3 = S_{\rm GN} e^{2 i \nu x}.
\label{3.5}
\end{equation}
The self-consistency condition for the GN spinor $\Psi$, Eq.~(\ref{3.1}), goes over into that of the isoNJL spinor $\Phi$, Eq.~(\ref{3.3}), in this process.  
Thus we get almost ``for free" an exact HF solution of the isoNJL model with all three chemical potentials different from zero. The resulting  
mean field ${\cal D}$, Eq.~(\ref{3.5}), can be fairly complicated, depending on the parameters, and is different from those of the one-flavor GN or NJL models.
Although $\Phi$ solves the HF equations for two standard NJL models, it does not satisfy the self-consistency
conditions of the two separate models, as one can easily check. This is in contrast to what happened in the special cases $\nu=\nu_5=0$ or $\mu=0$ above.
Therefore solution (\ref{3.5}) is genuine for the isoNJL model, even though the building blocks are borrowed from the one-flavor cases.

\begin{figure}
\begin{center}
\epsfig{file=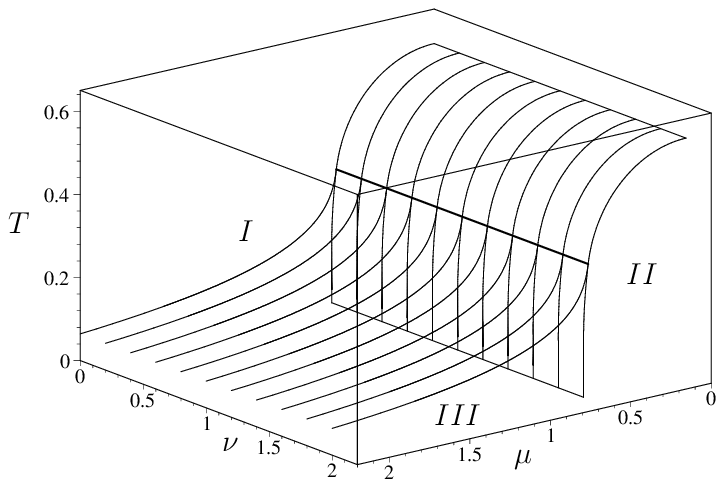,width=8cm,angle=0}
\caption{Full phase diagram of the massless isoNJL model. $I$) Chirally restored phase, $II$) chiral spiral with constant radius, $III$) chiral spiral with 
periodically modulated radius. There is no dependence on $\nu_5$.}
\label{fig2}
\end{center}
\end{figure}

Assuming that there is no better HF solution (where both ${\cal D}$ and ${\cal C}$ would have to be nonzero), we can now construct the phase diagram
of the isoNJL model in the chiral limit rather trivially, starting from Fig.~\ref{fig1}. Due to the factorization of the mean field,   
the 2nd order phase boundaries separating the symmetric, homogeneous and crystal phases of the GN model are just transported parallel in the $\nu$-direction, giving rise 
to the sheets of 2nd order phase transitions in ($\mu,\nu,T$)-space shown in Fig.~\ref{fig2}.  In region $I$, the mean field vanishes and chiral symmetry is restored. 
In region $II$, ${\cal D}$ has the form of the usual chiral spiral with fixed (temperature dependent) radius.
In region $III$, it is given by the product of the (finite temperature) GN kink crystal potential and the NJL factor $e^{2i\nu x}$. This represents a chiral spiral whose 
radius is modulated by the kink crystal shape, whereas its pitch depends only on the isospin chemical potential.  
Just like the axial chemical potential $\mu_5$ in the one-flavor NJL model \cite{L19}, the axial isospin chemical potential $\nu_5$ does not manifest itself at all in the phase diagram, 
although some observables will depend on it.
The thermodynamic potential of the isoNJL model is closely related to that of the GN model,
\begin{equation}
\left. \frac{{\cal V}_{\rm eff}(\mu,\nu,\nu_5,T)}{2 N_c} \right|_{\rm isoNJL} = \left. \frac{{\cal V}_{\rm eff}(\mu,T)}{N_c} \right|_{\rm GN}   - \frac{\nu^2+\nu_5^2}{2\pi}.
\label{3.6}
\end{equation}
The isospin density and the axial isospin density are spatially constant and temperature independent,
\begin{equation}
\rho_3 = \langle \psi^{\dagger} \tau_3 \psi \rangle =2 N_c \left( \frac{\nu}{\pi} \right), \quad   \rho_{3,5} = \langle \psi^{\dagger} \tau_3 \gamma_5 \psi \rangle  = 2 N_c \left( \frac{\nu_5}{\pi} \right).
\label{3.7}
\end{equation}
This is a well-known consequence of chiral symmetry and the axial anomaly. The fermion density on the other hand is periodic in $x$ and identical to that of the GN model
(up to an overall factor $2 N_c$ instead of $N_c$).

The simple dependence of the thermodynamic potential on $\nu,\nu_5$ has an important implication for the duality transformation discussed above, see Eqs.~(\ref{2.10},\ref{2.11}).
Since expression (\ref{3.6}) is invariant under exchange of $\nu$ and $-\nu_5$, exchanging ${\cal D}$ and ${\cal C}$ is equivalent to choosing a different vacuum point on the
vacuum manifold and has no observable consequences. Hence the question of whether there is a charged pion condensate is irrelevant here from the physics point of view.

The solution presented above is an exact, analytical HF solution for the most general chemical potentials and temperature with novel
crystalline mean fields as compared to single-flavor cases. What is still missing is a comparison of the thermodynamic potential 
with other HF solutions discussed in the literature. 
We already know that the present solution is the most stable one in the cases $\nu=\nu_5=0$ (GN model) or $\mu=0$ (NJL model). Here we should like to 
compare the thermodynamic potential to the homogeneous solution studied in Ref.~\cite{L15} if all three chemical potentials are different form zero. For
the sake of simplicity we restrict ourselves to $T=0$ where essentially everything can be done analytically.

\begin{figure}
\begin{center}
\epsfig{file=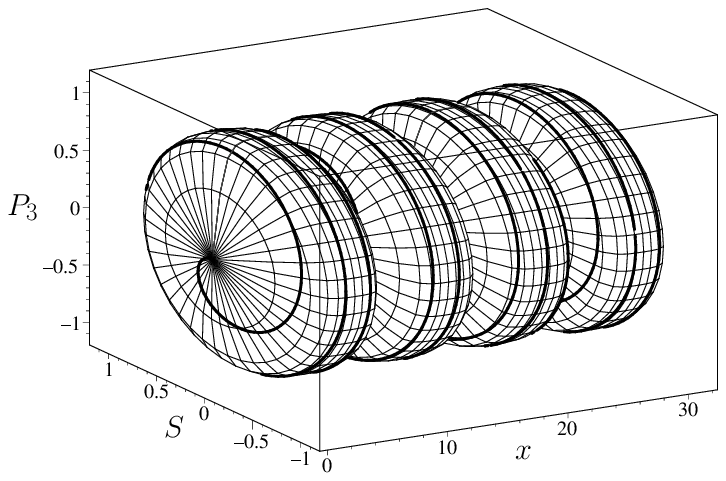,height=5cm,width=6cm,angle=0}
\caption{Example of order parameter in region $III$ of Fig.~\ref{fig2}. The radius of the chiral spiral is modulated by the kink-antikink crystal shape, see Eq.~(\ref{3.5}). Parameters: $\mu=0.637, \nu=1.5, T=0$.}
\label{fig3}
\end{center}
\end{figure}

\begin{figure}
\begin{center}
\epsfig{file=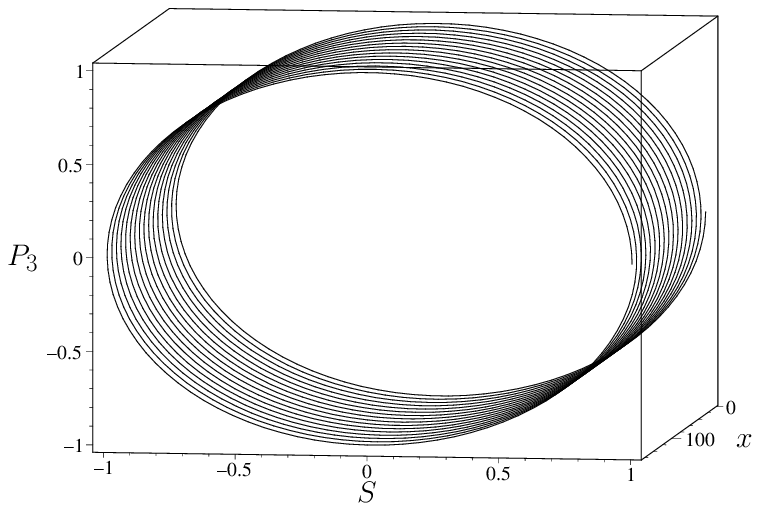,width=4.3cm,height=4cm,angle=0}\epsfig{file=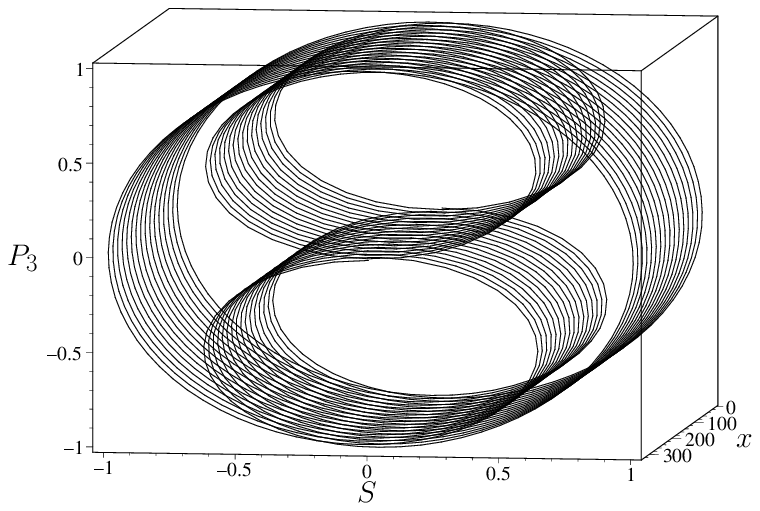,width=4.3cm,height=4cm,angle=0}\epsfig{file=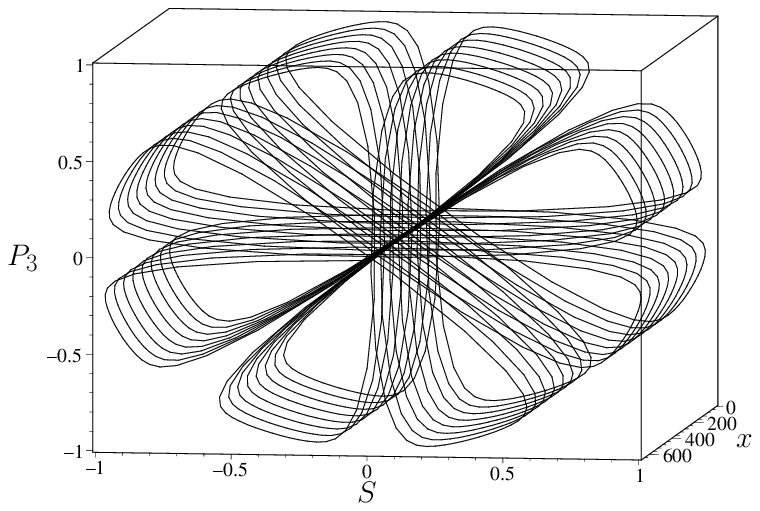,width=4.3cm,height=4cm,angle=0} 
\caption{Examples of order parameter for the case of commensurate spatial periods of $S_{\rm GN}$ and $e^{2 i \nu x}$ factors. The first plot is the standard chiral
spiral, the other two plots show additional possibilities with Lissajous character due to isospin imbalance. Parameters from left to right: i) $\mu<\mu_{\rm crit},\nu=0.306$, ii)
$\mu=0.637,\nu=0.306$, iii) $\mu=0.637,\nu=0.038$. The ratio of the NJL period to the GN period, Eq.~(\ref{3.10}), is 1/2 in ii) and 4 in iii). } 
\label{fig4}
\end{center}
\end{figure}

Let us first illustrate the mean field in the crystal phase at $T=0$ for a few cases. Using the explicit form of $S_{\rm GN}$ in terms of Jacobi elliptic functions \cite{L20}, 
the inhomogeneous HF potential of the isoNJL model is
\begin{equation}
{\cal D}(\mu,\nu,\nu_5, x) = \kappa \frac{{\rm sn}(\xi,\kappa) {\rm cn}(\xi, \kappa)}{{\rm dn}(\xi, \kappa)} e^{2i\nu x}, \quad \xi = \frac{x}{\kappa}.
\label{3.8}
\end{equation}
As pointed out above, it is independent of $\nu_5$, whereas the $\mu$ dependence is implicit in the following relation between the chemical potential $\mu$ and the elliptic
modulus $\kappa$,
\begin{equation}
\mu = \frac{2 {\bf E}(\kappa)}{\pi \kappa}.
\label{3.9}
\end{equation}
The spatial periods of the GN and NJL factors in ${\cal D}$ are
\begin{equation}
L_{\rm GN} = 2 \kappa {\bf K}(\kappa), \quad L_{\rm NJL} = \frac{\pi}{\nu}.
\label{3.10}
\end{equation}
For generic parameters $\mu, \nu$, these periods are incommensurate and the isoNJL mean field is not periodic. The best way of 
illustrating it is then the one shown in Fig.~\ref{fig3}. The fat line represents the space curve of the mean field, traced out on the surface
arising from rotating the kink-antikink crystal shape around the $x$ axis. If one chooses the two periods to be commensurate, one can  
generate chiral spirals with Lissajous character, see Fig.~\ref{fig4} where we compare two such curves with the standard NJL chiral spiral.
These examples should be sufficient to illustrate that the crystal structure of the isoNJL model is much richer than the one of the 
single-flavor NJL model.
In the incommensurate case, the plots analogous to Fig.~\ref{fig4} look rather chaotic and are not very illuminating.

In the GN model, the relationship between the grand canonical potential at $T=0$ and the chemical potential has been given in Sec.~4.3 of Ref.~\cite{L5},
\begin{equation}
\left. \frac{{\cal V}_{\rm eff}(\mu,T=0)}{N_c}\right|_{\rm GN}  =  \theta(\mu-\mu_{\rm crit}) \left(\frac{1}{2\pi} - \frac{1}{2\pi \kappa^2}\right), \quad \mu_{\rm crit}= \frac{2}{\pi}.
\label{3.11}
\end{equation}
The elliptic modulus $\kappa$ and the chemical potential $\mu$ are related by Eq.~(\ref{3.9}). 
The vacuum energy density has been subtracted as usual, so that the effective potential vanishes for $\mu \le \mu_{\rm crit}$. The critical chemical potential
$\mu_{\rm crit}$ corresponds to $\kappa=1$ (low density limit) and agrees with the baryon mass in the GN model \cite{L21} divided by $N_c$ (vacuum fermion mass = 1).
This result together with (\ref{3.6}) gives the following simple expression for the effective potential of the isoNJL model at $T=0$
\begin{equation}
\left. \frac{{\cal V}_{\rm eff}(\mu,\nu,\nu_5,T=0)}{2 N_c} \right|_{\rm isoNJL} = \theta(\mu-\mu_{\rm crit}) \left(\frac{1}{2\pi} - \frac{1}{2\pi \kappa^2}\right)  - \frac{\nu^2+\nu_5^2}{2\pi}.
\label{3.12}
\end{equation}
We can now compare our solution with the homogeneous one in the whole ($\mu,\nu,\nu_5$)-space. To this end we have repeated the calculation with constant 
mean field ${\cal D}$ (${\cal C}=0$) for the isoNJL model \cite{L15}. The result for the effective potential at $T=0$ can be cast into the form 
\begin{eqnarray}
\left. \frac{{\cal V}_{\rm eff}(\mu,\nu,\nu_5,T=0)}{2N_c}\right|_{\rm hom} &  = &  \min_{m} \Bigg\{  {\cal F}(m,\mu-\nu)+ {\cal F}(m,\nu-\mu) + {\cal F}(m,\mu+\nu) 
\nonumber \\ 
& & \left. +  {\cal F}(m,-\mu-\nu) +  \frac{1}{4 \pi}-\frac{m^2}{4\pi}(1-\ln m^2)-\frac{\nu_5^2}{2\pi}\right\}
\label{3.13}
\end{eqnarray}
where 
\begin{equation}
m=|{\cal D}|, \quad {\cal F}(m,y) = \theta(y-m)\frac{1}{4\pi} \left[ m^2 \ln \left( \frac{\sqrt{y^2-m^2}+y}{m}\right) - y \sqrt{y^2-m^2}\right].
\label{3.14}
\end{equation}
This is actually more complicated than (\ref{3.12}).
For constant ${\cal C}$ (${\cal D}=0$), one gets the same expression except that now $m=|{\cal C}|$ whereas  $\nu$ and $\nu_5$ are interchanged. As indicated in Eq.~(\ref{3.13}), 
the right-hand side still has to be minimized with respect to $m$, the physical fermion mass in matter. If the results for neutral and charged pion condensates are different,
one has to pick the lower one. This gives rise to a fairly complicated phase diagram, with regions of charged and neutral condensates as well as
chirally restored ones separated by first order phase transitions \cite{L15}. The origin of these complications are the step functions in (\ref{3.14}) which govern
when a certain branch of eigenvalues of $H$ crosses zero. We then compare the value in the minimum with our prediction (\ref{3.12}) in the relevant 
region of ($\mu,\nu,\nu_5$)-space (we looked at $0 <  \mu, \nu, \nu_5 < 2$). 

We find that the inhomogeneous solution is favored everywhere over the homogeneous one, at $T=0$. The only region in ($\mu,\nu,\nu_5$)-space where the two
calculations coincide is on the strip $\nu=0, \mu<\mu_{\rm crit}$, any $\nu_5$, where the new solution is also homogeneous. Since our result depends only on
$\nu^2+\nu_5^2$ as far as the isospin chemical potentials are concerned, there is no need to invoke duality. It makes no difference whether we work with
neutral or charged condensates. 

This comparison also gives a clue why the inhomogeneous potential is favored. In the GN model, the kink crystal potential is such that there is always a gap at the Fermi surface
(Peierls instability).
The unitary transformations applied later on to induce $\nu,\nu_5$ do not change this feature, moving the gapped fermion spectra rigidly up or down while keeping the occupation of each level fixed.
By contrast, the homogeneous calculation does not generate a gap at the Fermi surface, except in the vacuum.

The question still remains whether we are allowed to assume that either ${\cal D}$ or ${\cal C}$ vanish. At $\nu=\nu_5=0$ (isospin symmetric matter), a better solution 
with nonvanishing ${\cal C}$ and ${\cal D}$ can be ruled out thanks to the unbiased numerical investigation of Heinz {\em et al.} \cite{L12}. But what about isospin asymmetric systems?
Here we can only be certain that no better homogeneous solution exists, as shown by Khunjua {\em et al.} \cite{L15}. 
We believe that a better inhomogeneous HF solution with nonzero neutral and charged condensates is also unlikely. 
To this end, it is instructive to look at the chirally restored phase. The grand canonical potential density in the symmetric phase is, using (\ref{3.12},\ref{3.13}),
\begin{equation}
\left. \frac{{\cal V}_{\rm eff}(\mu,\nu,\nu_5,T=0)}{2N_c}\right|_{m=0}   = \frac{1}{4\pi} - \frac{\mu^2+ \nu^2 + \nu_5^2}{2\pi}  
\label{3.15}
\end{equation}
The constant term comes from the subtraction of the interacting vacuum energy density and would be absent in the free, massless theory.
This shows that the dependence of our full calculation on $\nu,\nu_5$ can hardly be improved, being the same as in the free, massless Fermi gas.
It is difficult to imagine that the interacting theory could require even less energy than the minimal amount dictated by kinematics and the Pauli principle, as
realized in a free, massless Fermi gas. 

\section{Massive isoNJL model -- Perturbative phase boundary sheet}
\label{sect4}

We now add a bare mass term to the Lagrangian (\ref{2.1}),
\begin{equation}
{\cal L} = \bar{\psi} ( i \partial \!\!\!/ - m_0)  \psi + \frac{G^2}{2} \left[ (\bar{\psi} \psi)^2+ (\bar{\psi} i \gamma_5 \vec{\tau}\psi)^2 \right] + \mu \psi^{\dagger} \psi
+ \nu \psi^{\dagger} \tau_3 \psi.
\label{4.1}
\end{equation}
The mass term breaks chiral symmetry explicitly and gives rise to a unique vacuum with $S=m, P_a=0$. Since the axial isospin charge is no longer conserved, we 
should not introduce an axial isospin chemical potential $\nu_5$ here. 
The gap equation (\ref{2.20}) is replaced by (using units where $m=1$ in the vacuum)
\begin{equation}
\frac{\pi}{2N_c G^2} = \gamma + \ln \Lambda, \quad \gamma= \frac{\pi m_0}{2 N_c G^2} = m_0 \ln \Lambda.
\label{4.2}
\end{equation} 
With the identification $2G^2=g^2$, the ``confinement parameter" $\gamma$ is the same as in the single-flavor model.
In case I of Sec.~\ref{sect2} ($\mu \neq 0, \nu=\nu_5=0$),
all the arguments leading to the GN model go through literally. We only need to replace $S_{\rm GN}$ by $S_{\rm GN}-m_0$ and ${\cal D}$ by ${\cal D}-m_0$ in the 
self-consistency relations (\ref{2.17}) and (\ref{2.19}), respectively.
This implies that the phase diagram of the massive isoNJL model at finite ($\mu,T$) is the same as that of the massive GN model \cite{L6}.
Unlike in the chiral limit, there is as yet no independent confirmation of this result.
In the second special case ($\mu=0, \nu \neq 0$), the arguments in favor of the NJL phase diagram can be repeated once again,
except that the axial chemical potential should be omitted. Hence the phase diagram in ($\nu, \gamma, T$)-space of the massive isoNJL model should be identical to 
the phase diagram in ($\mu, \gamma, T$)-space of the massive one-flavor NJL model \cite{L7}.
Based on the results of the earlier studies of massive GN and NJL models, we may then replace
Fig.~\ref{fig1} in the chiral limit by Fig.~{\ref{fig5} for the massive isoNJL model, without any additional effort.
However, it is not obvious what replaces Fig.~\ref{fig2} in the massive case. 
Since chiral symmetry is explicitly broken, the trick with the local unitary transformation used above is not available
anymore. We have to resort to other methods to construct the full phase diagram in the bulk of ($\mu,\nu,T$)-space.

\begin{figure}
\begin{center}
\epsfig{file=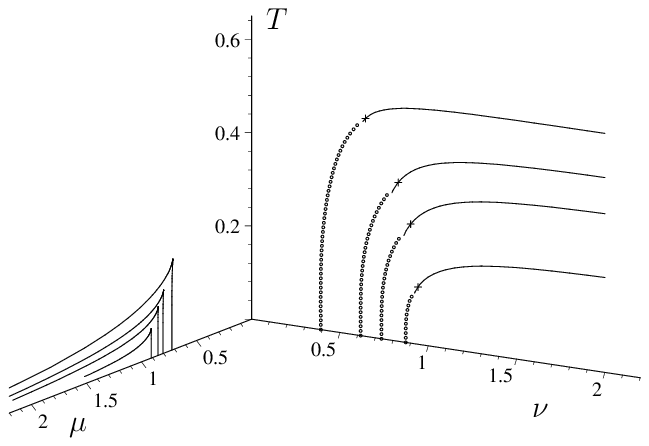,width=8cm,angle=0}
\caption{Phase diagram of the massive isoNJL model in the ($\mu,T$) and ($\nu,T$) planes, generalizing Fig.~\ref{fig1} of the massless model. The curves in the
$\nu=0$ plane can be taken over from the massive GN model, the curves in the $\mu=0$ plane from the massive NJL model. Only curves with $\gamma=0.1,0.3,0.5,1.0$
(from top to bottom) are shown.}
\label{fig5}
\end{center}
\end{figure}

Before continuing, it is worthwhile to recall the character of the critical lines drawn in Fig.~\ref{fig5} and taken from Refs.~\cite{L6,L7}. 
These lines separate the homogeneous from the inhomogeneous phases. The curves from both
single-flavor models (GN and NJL) exhibit a tricritical point, dividing them into a ``horizontal" and a ``vertical" section 
(the latter connects the tricritical point with the $T=0$ axis). Along the ``horizontal" phase boundaries, a second order phase transition occurs where the mean field vanishes 
in a continuous fashion. Hence these phase boundaries can be determined using perturbation theory in the mean field. By contrast, the ``vertical"
phase boundaries are nonperturbative. In the case of the GN model [($\mu,T$)-plane in Fig.~\ref{fig5}], they are characterized by the instability of the 
homogeneous system toward formation of a single baryon. The phase transition is 2nd order but nonperturbative. 
In the case of the NJL model [($\nu,T$)-plane in Fig.~\ref{fig5}], the phase transition is discontinuous and of first order, i.e., there are two competing local minima of the same depth
in the effective potential along the phase boundary. In order to complete the phase diagram in the whole ($\mu,\nu,T$)-space, we shall assume that this
division into perturbative and nonperturbative phase boundaries makes sense everywhere. In the present section, we determine 
the ``horizontal", perturbative phase boundary sheet of the isoNJL model, using a simple stability analysis. 

Since the problem at hand is closely related to one-flavor NJL models for each isospin component, we can take over some technicalities from the study of the massive NJL model \cite{L7}.
Let us briefly recall the approach used there. Starting point was the HF Hamiltonian divided up into
\begin{equation}
H = H_0 +V
\label{4.3}
\end{equation}
where
\begin{eqnarray}
H_0 & = &  - \gamma_5 i \partial_x + \gamma^0 m,
\nonumber \\
V & = & \gamma^0 2 S_1 \cos (2 q x) - i \gamma^1 2 P_1 \sin(2qx).
\label{4.5}
\end{eqnarray} 
The wave number $q$ where the instability occurs was denoted by $k_f$ in Ref.~\cite{L7}. For arguments why this ansatz is general enough to find the exact phase boundary, 
we refer to that paper. The change in the grand canonical potential due to $V$ was determined in 2nd order almost degenerate perturbation theory (ADPT) with the result
\begin{eqnarray}
\delta \Psi_{\rm NJL}(\mu,T,S_1,P_1,q) & = & \frac{E_q^2 S_1^2+q^2P_1^2}{\pi} \int_0^{\infty}\!\!\!\!\!\!\!\!\!\!\!- dp \frac{1}{E(p^2-q^2)}
\left( \frac{1}{e^{\beta (E-\mu)}+1} + \frac{1}{e^{\beta (E+\mu)}+1} \right)
\nonumber \\
&  & + \frac{2qS_1 P_1}{\pi} \int_0^{\infty}\!\!\!\!\!\!\!\!\!\!\!- dp \frac{1}{p^2-q^2} 
\left( \frac{1}{e^{\beta (E-\mu)}+1} - \frac{1}{e^{\beta (E+\mu)}+1} \right)
\nonumber \\
& & + \frac{S_1^2+P_1^2}{\pi} \frac{\gamma}{m} - \frac{E_q^2 S_1^2+q^2 P_1^2}{2\pi q E_q} \ln \left( \frac{E_q-q}{E_q+q} \right)
\label{4.6}
\end{eqnarray}
where
\begin{equation}
E=\sqrt{m^2+p^2},\quad E_q=\sqrt{m^2+q^2}.
\label{4.7}
\end{equation}
The only remnant of ADPT in this expression is the principal value prescription for integrating through the singularity at $p=q$.
Requiring that the grand potential be stationary with respect to $m,S_1,P_1,q$ then leads to the conditions
\begin{equation}
{\rm det} {\cal M} = 0, \quad \partial_q {\rm det} {\cal M} = 0
\label{4.8}
\end{equation} 
with ${\cal M}$ the ``Hessian"  matrix
\begin{equation}
{\cal M} = \left( \begin{array}{cc} \partial_{S_1}^2 (\delta \Psi)  & \partial_{S_1}\partial_{P_1} (\delta \Psi) \\   \partial_{P_1}\partial_{S_1} (\delta \Psi) & \partial_{P_1}^2  (\delta \Psi) \end{array} \right)
\label{4.9}
\end{equation} 
The mass $m$ appearing in this expression is the mass which minimizes the homogeneous grand canonical potential at the same
temperature and chemical potential. It is now straightforward to adapt this computation to the massive isoNJL model. All we have to do is replace $\delta \Psi_{\rm NJL}$ by
\begin{equation}
\delta \Psi_{\rm isoNJL}(\mu,\nu,T,S_1,P_1,q) = \frac{1}{2} \left[ \delta \Psi_{\rm NJL}(\mu+\nu,T,S_1,P_1,q) + \delta \Psi_{\rm NJL}(\mu-\nu,T,S_1,-P_1,q) \right]
\label{4.10}
\end{equation}
The chemical potentials $\mu \pm \nu$ and the fact that the sign of $P_1$ had to be reversed in the isospin down contribution can be read off from Eq.~(\ref{2.12}), remembering 
that $\nu_5=0$ in the massive model.

In order to locate the perturbative phase boundary, we work at a fixed $\gamma$ and scan the ($\mu,\nu$)-plane on a regular grid. At each grid point, we
vary the temperature until the determinant and its derivative vanish simultaneously, see Eq.~(\ref{4.8}). We have first tested our procedure in the 
chiral limit where the answer is already known (Fig.~\ref{fig2}). In this limit, the perturbative sheet actually consists of two distinct parts separated by 
a tricritical line. The part at $\mu<\mu_{\rm crit}$ separates the symmetric phase from a standard chiral spiral crystal (constant radius), whereas the part at 
$\mu>\mu_{\rm crit}$ separates the symmetric phase from the novel chiral spiral crystal (periodically modulated radius). We find that the perturbative calculation 
maps out both parts accurately. As a matter of fact, by looking at the plot of ${\rm det}({\cal M})$ versus $q$ on the phase boundary, one gets additional information
about the character of the instability. This is shown in Fig.~\ref{fig6}. At $\mu<\mu_{\rm crit}$, we see only one critical wave number, $q=\nu$.
At $\mu>\mu_{\rm crit}$ two critical wave numbers appear simultaneously, $q=\mu + \nu$ and $q=|\mu-\nu|$. This is due to the factorization
of $\Delta$ into a chiral spiral factor and a soliton crystal profile, the latter also reducing to a plane wave at the phase boundary. At $\mu=\mu_{\rm crit}$,
we find that the 2nd derivative of the determinant  vanishes as well.

\begin{figure}
\begin{center}
\epsfig{file=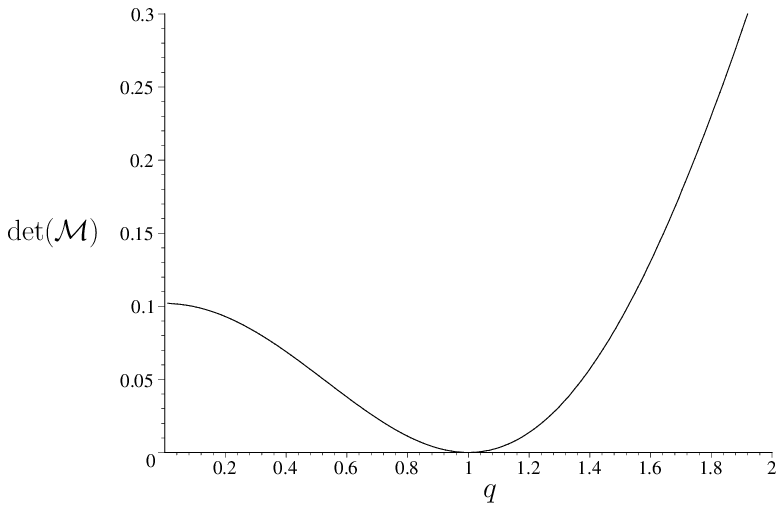,width=5.5cm,angle=0}\epsfig{file=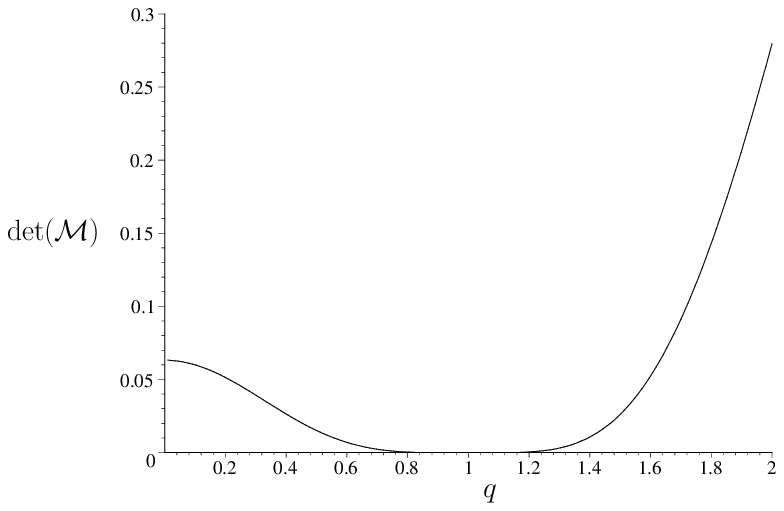,width=5.5cm,angle=0}\epsfig{file=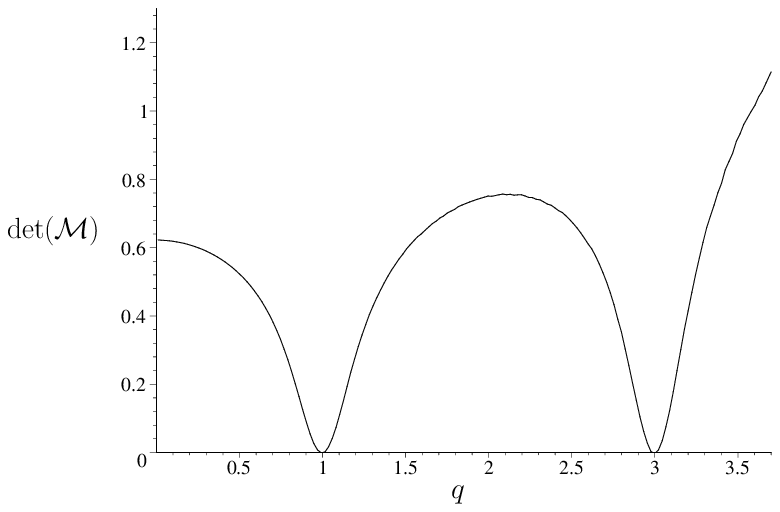,width=5.5cm,angle=0}
\caption{Examples of plots of det(${\cal M}$) right on the phase boundary sheet, as they are found in the massless isoNJL model. From left to right: 
i) Boundary between phases $I$ and $II$ in Fig.~\ref{fig2}, $\mu=0.3, \nu=1$, ii) Tricritical line where phases $I,II,III$ meet, $\mu=0.608,\nu=1$, iii)
boundary between phases $I$ and $III$, $\mu=2, \nu=1$.}  
\label{fig6}
\end{center}
\end{figure}

\begin{figure}
\begin{center}
\epsfig{file=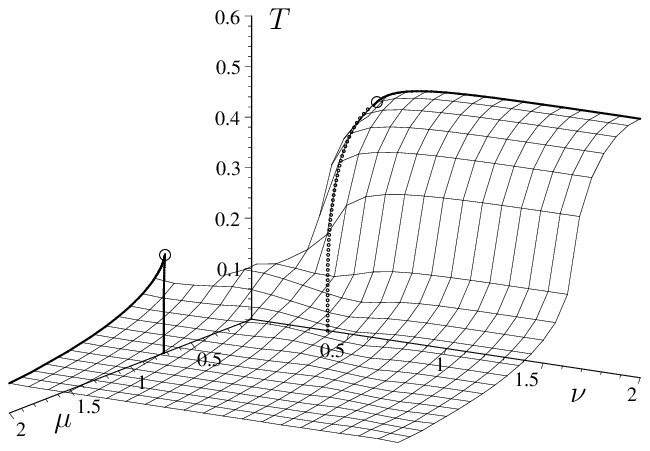,width=8cm,angle=0}
\caption{Perturbative phase boundary between crystal phase and massive Fermi gas in the massive isoNJL model. The result shown was obtained at $\gamma=0.1$,
corresponding to the uppermost curves in Fig.~\ref{fig5}.}
\label{fig7}
\end{center}
\end{figure}

We have performed such perturbative calculations of the phase boundary for the massive isoNJL model at $\gamma=0.1...0.9$ in steps of $0.1$. 
By way of example, Fig.~\ref{fig7} shows how the horizontal critical curves of the GN model (at $\nu=0$) are connected to those of the NJL model (at $\mu=0$)
by a smooth phase boundary sheet. As expected from the GN and NJL models, no perturbative phase boundary
could be found in a region close to the origin, below the crystal phases of the two models. There must be a homogeneous region extending down to 
the $T=0$ plane, presumably separated from the crystal phase by a first order transition. This part of the phase diagram cannot be captured by a 
perturbative stability analysis but would require a full numerical HF calculation. Unfortunately, plots of det(${\cal M}$) are not as simple as in the chiral 
limit, Fig.~\ref{fig6}, so that we do not get any clue as to the crystal structure below the sheet, or the position of a possible tricritical line.

In order to illustrate the dependence of the phase diagram on the parameter $\gamma$, we show all of our results in Fig.~\ref{fig8}.
In all cases considered, we find that the novel perturbative phase boundary sheet in the bulk of ($\mu,\nu,T$) space interpolates smoothly between the 
known GN and NJL results on the boundaries. With increasing bare fermion mass, the critical temperature where the crystal phase ends decreases.
One would expect that the inhomogeneous phase survives at $T=0$ and sufficiently large chemical potentials. The main qualitative difference between 
massless and massive isoNJL models is the ``wound" around the origin where a homogeneous island is expected. Judging from the GN and NJL models,
there should be a nonperturbative, ``vertical"
phase boundary connecting the perturbative sheet with the $T=0$ plane. To determine it would require a full numerical HF calculation as described 
in Ref.~\cite{L7} for the massive NJL model.

The only other conspicuous feature of the perturbative phase boundaries is a slight but systematic enhancement in the direction of the diagonal $\mu=\nu$.
We have no explanation for this phenomenon, so that we have to await the results of the complete HF calculation inside the crystal phase which is 
underway. 

\begin{figure}
\begin{center}
\epsfig{file=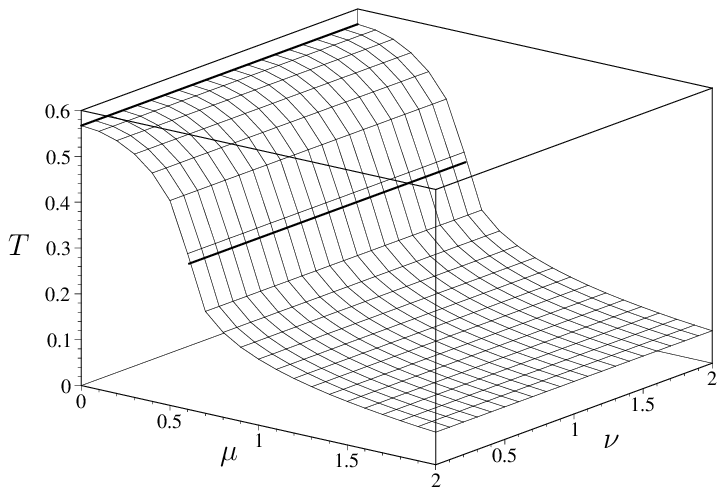,width=6cm,angle=0}\quad \epsfig{file=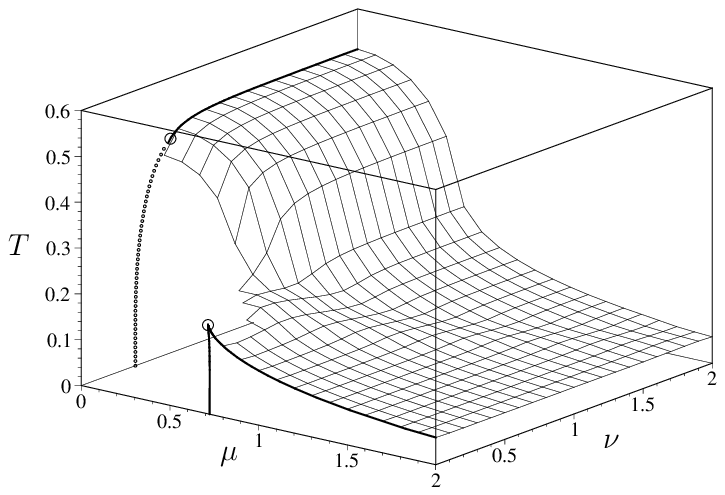,width=6cm,angle=0}
\epsfig{file=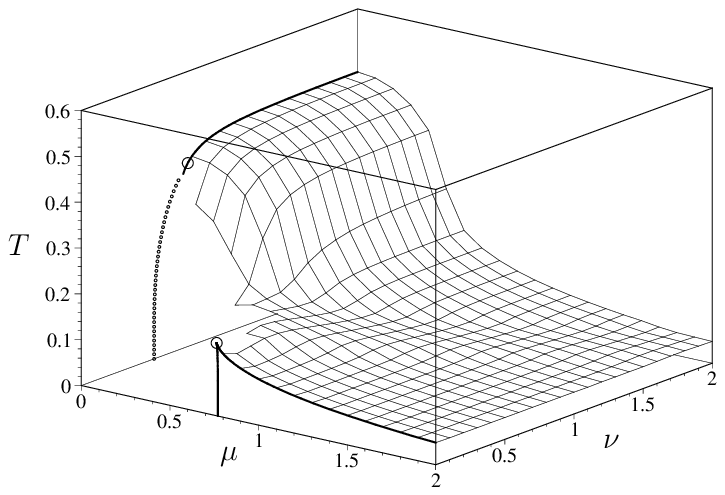,width=6cm,angle=0}\quad \epsfig{file=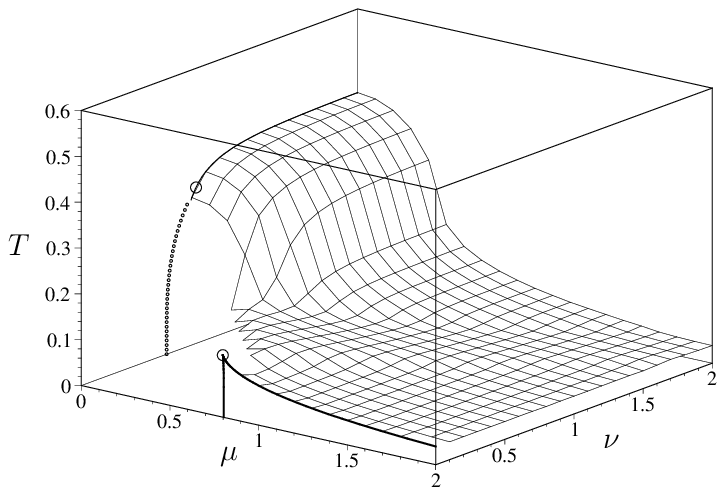,width=6cm,angle=0}
\epsfig{file=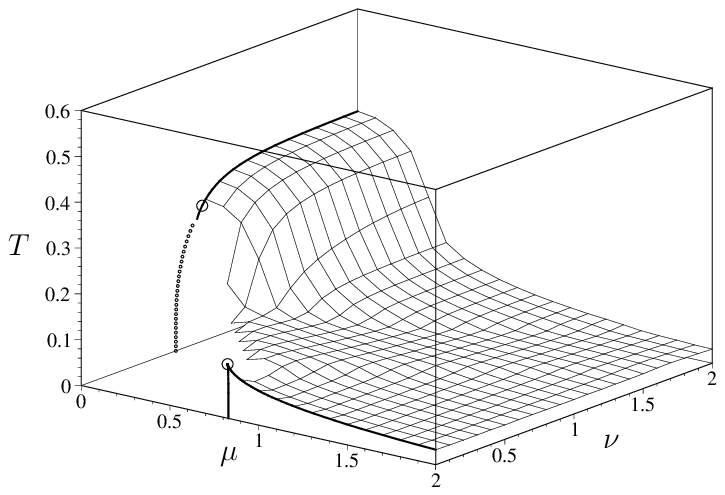,width=6cm,angle=0}\quad \epsfig{file=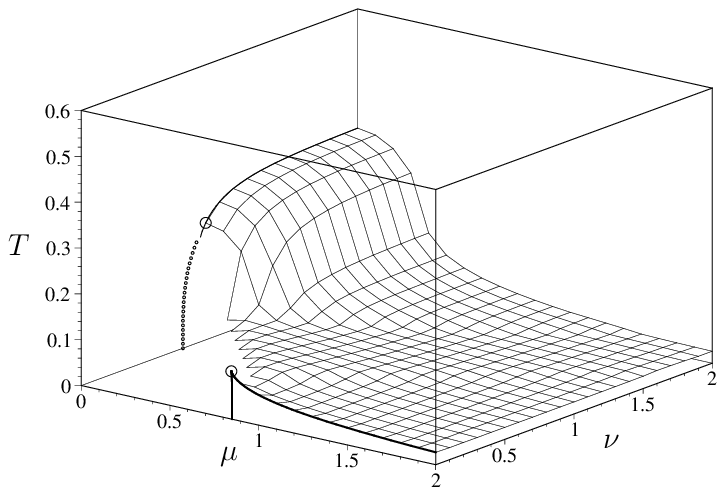,width=6cm,angle=0}
\epsfig{file=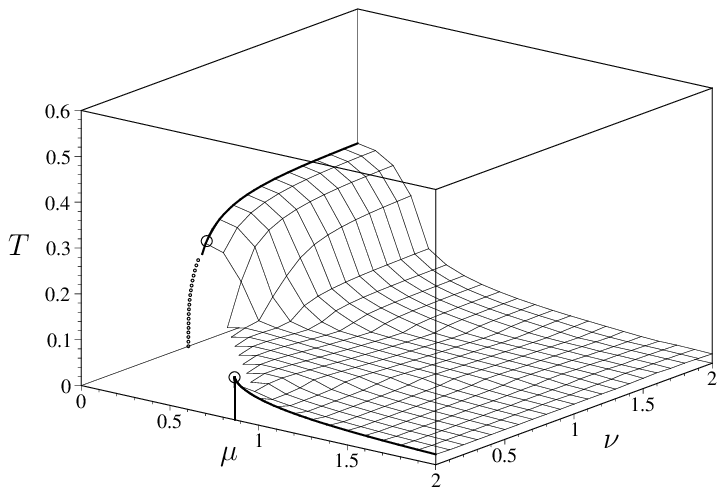,width=6cm,angle=0}\quad \epsfig{file=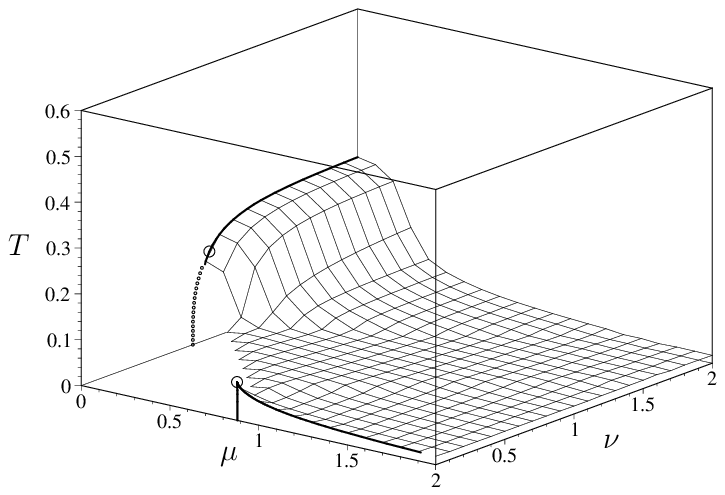,width=6cm,angle=0}
\epsfig{file=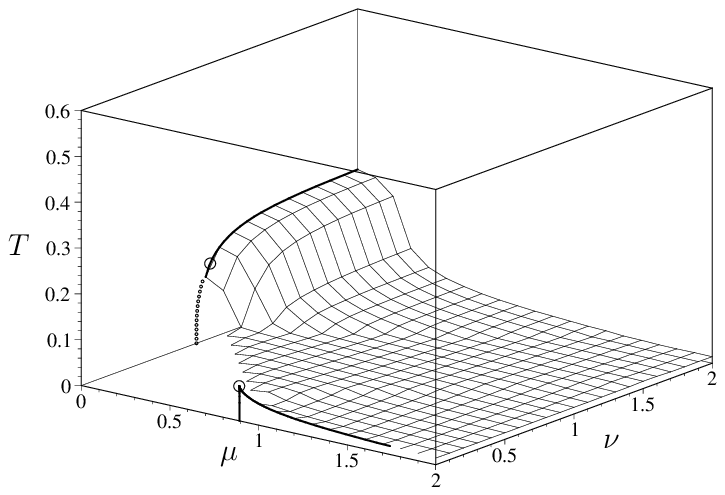,width=6cm,angle=0}\quad \epsfig{file=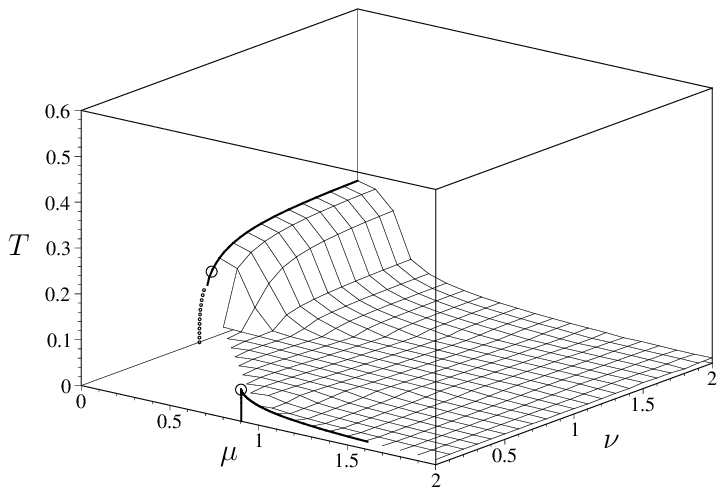,width=6cm,angle=0}
\caption{Perturbative phase boundary sheets of massive isoNJL model for $\gamma=0...0.9$ in steps of 0.1. Note the systematic slight enhancement 
along the diagonal of the ($\mu,\nu$) plane. A homogeneous region down to $T=0$ is expected in the hole visible near the origin. The 
nonperturbative phase boundary delimiting this hole is not yet known, except on the $\mu=0$ and $\nu=0$ planes.}
\label{fig8}
\end{center}
\end{figure}

\section{Summary and conclusions}
\label{sect5}

In this paper, we have studied the phase diagram of the isoNJL model in 1+1 dimensions. In the chiral limit, we have arrived at the full phase diagram
as a function of temperature and all three chemical potentials corresponding to fermion density, iospin density and chiral isospin density, using 
only analytical tools. This would not have been possible without prior analytical knowledge of the phase diagrams of the one-flavor GN and NJL
models. The chiral isospin chemical potential does not show up at all in the phase diagram and the 3-dimensional ($\mu,\nu,T$)-space is sufficient
to exhibit the phase structure. In the ($\mu,T$)-plane, we recover the GN phase diagram, as noticed previously by others using numerical computations.
In the ($\nu,T$)-plane, the phase diagram is identical to that of the NJL model in the ($\mu,T$)-plane. The interesting question is then how the system 
interpolates in the bulk of ($\mu,\nu,T$)-space between these two well-known, qualitatively different crystal phases. The answer is a factorization 
of the order parameter into those of the GN and the NJL model. The resulting phase diagram can be generated by a parallel transport of the GN phase diagram
in the direction of the $\nu$ axis. In spite of this simple construction for which the chiral anomaly is instrumental, the order parameter can look quite complicated.
It would have been difficult to understand it using only numerical methods.

The basic assumption behind these results is that one can set the charged condensate (${\cal C}=P_2-iP_1$) equal to zero. We have given heuristic
arguments and empirical evidence that this is reasonable, at least in the chiral limit. The most convincing argument in our opinion is the fact that at zero
temperature we can make sure that there is always a gap at the Fermi surface, in accordance with the Peierls instability in condensed matter physics.
This cannot be achieved with homogeneous phases only. Nevertheless one should try to rule out a more complicated order parameter with nonvanishing
neutral and charged condensates in a more rigorous fashion in future work.

Turning to the massive isoNJL model, we first pointed out that in the ($\mu,T$) and ($\nu,T$)-planes, the phase diagram is again identical to that of the 
massive GN and massive NJL models, respectively. These phase diagrams in turn are known analytically (GN) or at least numerically (NJL) in great detail.
Both of these simpler models feature a perturbative 2nd order phase boundary separating the crystal from the homogeneous phase beyond the tricritical point. 
This type of phase boundary can be determined in a rather straightforward way by a stability analysis without the need to do a full, self-consistent HF calculation.
This has incited us to construct the perturbative phase boundary sheet for the isoNJL model in the bulk of ($\mu,\nu,T$)-space as well. We have constructed smooth
surfaces interpolating between the corresponding curves of the GN and NJL model which already give a fairly complete impression of what the full phase diagram
will look like. The only region which cannot be understood in this manner is the (homogeneous) ``hole" around $\mu=0,\nu=0$ characteristic for the massive theory.
Moreover, there could well be further phase boundaries inside the inhomogeneous region separating different types of crystal, as is already the case in the chiral limit.
In order to answer these questions, a complete HF calculation with a nontrivial numerical effort is unavoidable. Such calculations are underway, and the results will 
be presented elsewhere. We also intend to relax the assumption that either the charged or the neutral condensate vanishes. While this has been crucial for the present 
analytical and (perturbative) numerical work, we have to admit that the arguments in its favor are somewhat weaker in the massive theory than in the chiral limit.

\end{document}